# λ-persistant CSMA: a radio-channel access protocol[*]

Przemysław Błaśkiewicz, Jacek Cichoń, Jakub Lemiesz,
Mirosław Kutyłowski, Marcin Zawada
Wroclaw University of Technology, Wroclaw, Poland
Email: {firstname.lastname}@pwr.edu.pl

Szymon Stefański, Krzysztof Chrobak
DATAX Sp. z o.o., Wroclaw
Email: szymon.stefanski@datax.pl, kchrobak@datax.pl

*Abstract*—This paper presents an algorithm that improves channel-access statistics for wireless medium. The proposed modification of the standard CSMA algorithm is analytically shown to yield better results and simulation results are given to support this claim.

## I. INTRODUCTION

One of the fundamental problems related to communication in wireless networks using a shared channel is the method of coordinated access of different stations to the transmission mdium. The lack of coorination leads to simultaneous communications and, consequently, message loss or need to re-transmit.

This problem is particularly difficult for spontaneous, ad-hoc networks, such as these offered by 802.15.4 standard [1] and ad-hoc WLANs based on IEEE 802.11 specification [2]. The use of a radio channel allows for spontaneous organization of communication without the need for permanent infrastructure and devices can connect and disconnect without prior announcement.

This paper presents two improvements of a distributed algorithm for medium access used in wireless networks, namely *p*-presistent CSMA. We provide analytical derivation of the probability of success for sending messages by stations. Our analysis shows that the modification performs significantly better with respect to channel availability and average waiting time than the basic algorithm. We also provide simulation results in two different load generation scenarios: full buffer, when stations always have buffered messages to be sent (heavily loaded network), and a poisson-distributed, simulating standard operation.

Sections 2 and 3 provide a short insight into problems of channel-access methods. The new protocols are presented in Section 4, followed by simulation results in Section 5 and conclusion in Section 6.

## II. MEDIUM ACCESS PROTOCOLS

In stationary wireless networks (such as cellular networks, for example), where particular stations (eg. BTS statsions) are responsible for management of the radio, one can consider TDMA (time-division multiple access) algorithms for accessing the medium, where each station has its own assigned moments in time when it is allowed to transmit. On the other hand, however, while considering ad-hoc, no-infrastructure wireless networks, it is effective to think of an *ansynchronous* model, where the stations have no external mechanism for coordination of radio channel usage. It is the role of MAC (Medium ACcess) protocols to provide such service. There is a number of approaches to this problem. The first solution wat that offered by ALOHA [3], [4] and its synchronous version - slotted ALOHA [5], where a station attempting to transmit, is able to detect if other stations are using the medium. When the medium is free, the station transmits, and *backs-off* for a random period of time otherwise.

### A. Collision avoidance in CSMA

In this paper, as well as in many other approaches, we focus on methods where a station can (as in ALOHA) determine if the medium is being used by means of *carrier sensing* – CS. All CSMA (carrier sensing multiple access) methods require that a station, prior to sending its message, listens to the medium to determine if there is any transmission going on. This verification can be made on physical layer – by determining for example S/N ratio (see eg. [6]) or can utilise upper layers of transmission protocol (eg. by determining the preamble of a packet).

Typically, one distinguishes two cases for CSMA: a more advanced collision detection (CSMA/CD) variant, where a station can determine if its transmission colided with some other, and collision avoidance (CSMA/CA) – where the collision can be avoided (in expectation or by means of protocol) but, if it happens – cannot be detected. In this paper we focus on the latter case.

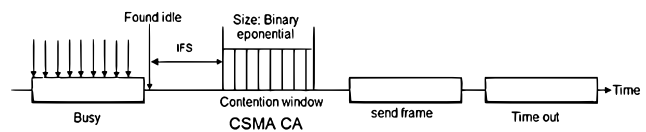

Fig. 1. CSMA/CA

CSMA/CA avoids the collisions using three basic techniques (cf. Fig. 1.

[*]Supported by National Centre for Research and Development project no. PBS1/A3/6/2012

i) *Interframe space (IFS)*. Whenever the channel is found idle, the station does not transmit immediately. It waits for a period of time called interframe space (IFS). When channel is sensed to be idle, it may be possible that some distant station may have already started transmitting and the signal of that station has not yet reached other stations. Therefore, the purpose of IFS time is to allow this transmitted signal to reach other stations. If after IFS time the channel is still idle, the station can send, but it still needs to wait a time equal to contention time. IFS variable can also be used to define the priority of a station or a frame.
ii) *Contention window*. Contention window is an amount of time divided into slots. A station that is ready to send chooses a random number of slots as its waiting time. The number of slots in the window changes according to the binary exponential back-off strategy. It means that it is a set of one slot the first time a station tries to send, and then doubles each time the station cannot detect an idle channel after the IFS time. This is very similar to the $p$-persistent method (see below) except that a random outcome defines the number of slots taken by the waiting station. In contention window the station needs to sense the channel after each time slot. If the station finds the channel busy, it does not restart the process. It just stops the timer and restarts it when the channel is sensed as idle.
iii) *Acknowledgements*. Despite all the precautions, collisions may occur and destroy the data. A positive acknowledgment and a time-out timer can help guarantee that receiver has received the frame.

The other factor differentiating protocols in that family is the way the station behaves after testing the channel:
i) *non-persistent CSMA* – when the channel is free, the station transmits; otherwise, the station backs-off for a period of time before re-initiating the whole procedure;
ii) *$p$-persistent CSMA* – when the channel is free, the station transmits with a certain probability $p$, and with $(1-p)$ it re-initializes the procedure after sime time $\alpha$; when the channel is busy, the station actively waits for the end of the transmission and re-initializes the procedure again.

An important case of the latter behaviour is when $p = 1$, i.e., when the station waits till the end of the ongoing transmission to immediately start it own. Therefore, when two stations start execution of the protocol during some transmission, their collision must happen. On the other hand, the end of previous transmission can be used as a synchronising moment for the network, and itroducing $p < 1$ allows an effective dispersion of the stations that have piled up awaiting for the end of the transmission. In our work we leverage these two observations and provide a distribution for $p$ that reduces the probability of collision, so that the network throughput is maximised.

## III. RELATED WORK

In the literature on CSMA methods, there are two different areas of research. The first focuses on mechanisms of determining such *squelch* levels, that any signal above the threshold can be viewed as a tranmission. In [7] the authors present a theoretical model allowing calculation of the threshold in regular networks, given their desired average throughput and with respect to their topology and minimal SINR (which, by Shannon's theorem, indicates the available transmission bandwith). This model was further extended in [8] to include the overhead caused by any MAC imposed on the network: the throughput must decrease whenever stations compete for access to the medium (and, naturally, only one among many can transmit). At the same time, the paper [9] shows that using CS methodology allows reaching throughout close to the theoretical limit, conditioned on some flexibility of stations' bitrate.

The other direction of research is that of methods for solving congestions of traffic that builds up while other transmissions are in progress (see Fig.2). The research focuses on determining the moment ($\eta$) of transmission start such that bandwith usage is maximised on the one hand, and collision is avoided on the other. The paper [10] presents a mechanism for dynamical calculation of back-off time in dynamic WLAN networks, which allows the near-optimal overall network throughput. In [11], the same approach is extended to multi-hop networks. Another idea is that of [12], where a station, with some constant probability verifies if the channel is free and then transmits a short probing message (a version of software handshake protocol). A similar mechanism is employed in [13], bearing the difference that the probability is conditioned on the number of stations in the network and the probing message is sent regardless of the channel state. A family of protocols is presented in [14] offering optimal probabilities for a range of different parameters of the network (number of stations, average delays, etc.). This paper is a continuation of work presented in [14] and offers improvement for solving congestion in $p$-persistent CSMA.

### A. Cai-Lu-Wang Protocol

In this paper we introduce improvements to the Cai-Lu-Wang protocol [12]. It depends on parameters $p \in [0, 1]$ and $T \geq 0$. At the beginning of a time *slot*, each station chooses $\xi \in [0, 1]$ uniformly at random and $\eta \in [0, T]$ also uniformly at random. If $\xi < p$, then at moment $T_0 + \eta$ the station monitors the channel. If the station does not detect carrier signal in the shared channel, then it starts transmitting a message of length $\delta$. * This message is received by the coordinator at moment not later than $T_0 + T + \delta + \lambda$. At this moment the coordinator may recognize whether a single station has transmitted. If it is so, then the coordinator sends SUCCESS. In the opposite case it transmits CONTINUE. These messages will be received by all stations at moment $T_0 + T + 2(\delta + \lambda)$. Therefore, we see, that the total length of a single slot is $T + 2(\delta + \lambda)$. Notice that if $T = 0$, then this algorithm reduces to the Nakano-Olariu algorithm.

---
*In fact, we present here a slightly modified algorithm in order expose the real difference with Nakano-Olariu method.

The calculations of the probability of success of one such trial is based on the following classical theorem:

**Theorem 1.** *Suppose that we randomly and independently choose random numbers $X_1, \ldots, X_n$ from $[0,1)$ according to the distribution $f$ with a cumulative density function $F$. Let $X_{1:n} \leq \ldots \leq X_{n:n}$ denote the sequence $X_1, \ldots, X_n$ after sorting it according to the current values. Then*

$$\Pr(X_{2:n} - X_{1:n} > \lambda) = n \int_0^{1-\lambda} f(x)(1 - F(x+\lambda))^{n-1} dx \ .$$

For the uniform distribution ($f \equiv 1$) on the interval $[0,1)$ this theorem gives

$$\Pr(X_{2:n} - X_{1:n} > \lambda) = (1-\lambda)^n \ .$$

Let $G_k$ denote the event that precisely $k$ stations in a given slot decide to transmit. Then for $k > 1$ we have

$$\Pr[\text{Success}|G_k] = (1 - \lambda/T)^k \ .$$

Notice that $\Pr[\text{Success}|G_1] = 1$. Therefore, if $T \geq \lambda$, then

$$\Pr[\text{Success}] = \sum_{k=1}^N \Pr[\text{Success}|G_k] \cdot \Pr[G_k] =$$

$$Np(1-p)^{N-1} + \sum_{k=2}^N (1-\lambda/T)^k \binom{N}{k} p^k (1-p)^{n-k} \ .$$

Notice also that if $T < \lambda$, then

$$\Pr[\text{Success}] = Np(1-p)^{N-1} \ .$$

Let $\text{CLW}_{p,T}$ be a random variable denoting the time necessary to choose a leader in this algorithm. The above remarks allows as to write a formula for $\text{E}[\text{CLW}_{p,T}]$ (see [12] for details). The behavior of this algorithm depends on a proper setting of parameters $p$ and $T$ for given parameters $\lambda, \delta$ and $n$. Let us remark that no analytical formula for the optimal choice of parameters is known.

Cai Lu and Wang observed in [12] that a slightly better run-time properties are achieved by protocols based on the probability distributions of the form $F(x) = x^\alpha$. We checked that solutions based on the distribution

$$F(x) = (e^{\alpha x^\beta} - 1)(e^\alpha - 1) \ ,$$

where $x \in [0,1]$ and $\alpha, \beta > 0$, have slightly better properties than the solution based on the distributions of the form $F(x) = x^\alpha$ (see [14]). However, we shall see that solutions based on discrete probabilities are even better.

In [15] we have proposed a modification of this protocol which works under a stronger assumption that a station (which has its receiver on) can recognize whether a collision has occurred during the current transmission. This algorithm has also better statistical properties than the original Cai-Lu-Wang solution.

## IV. NEW PROTOCOL

This part of the paper presents new medium access protocols leveraging leader election protocols from [12] and [14]. These algorithms utilise standard CS mechanism and improve over $p$-presistent CSMA.

### A. Protocol using order statistics

The first protocol modifies the $p$-presistend CSMA. Specifically, we propose a change in the part of the protocol executed when a station awaits for the end of the current transmission. The original behaviour would be here to transmit with probability $p$ and redo the transmission attempt after time $\lambda$ with probability $1-p$. In our version, all stations wishing to transmit wait until the channel becomes free (time $T_0$ in Fig. 2) and choose independently at random values $\xi \in [0,1]$ and $\eta \in [0,T]$, for some $T \geq 0$. Let $p \in [0,1]$, then if $\xi < p$ then at time $T_0 + \eta$ the station performs CS and starts its transmission when detects a free channel. If the channel is busy, it awaits the end of transmission and again chooses $\xi$ and $\eta$. This protocol is presented as Algorithm 1.

---

**Algorithm 1** Improved $p$-presistent CSMA

**Entry condition:**
1: there is an ongoing transmission

TX ATTEMPT:
1: **repeat** listen to channel
2: **until** transmission ends
3: $T_0 \leftarrow$ current moment
4: $\xi \leftarrow [0,1]$ ▷ uniformly, independently
5: $\eta \leftarrow [0,T], T \geq 0$ ▷ uniformly, independently
6: **if** $\xi < p$ **then**
7:     at time $T_0 + \eta$ perform CS
8:     **if** no carrier detected **then**
9:        transmit
10:     **else**
11:        redo TX ATTEMPT
12:     **end if**
13: **else**
14:     redo TX ATTEMPT
15: **end if**

---

### B. Protocol using constant delay and optimal non-uniform distribution

This version, similarly to the classical $p$-presistent SMA, the stations wait until the end of transmission. Then, all listening stations attempt to transmit at a number of *different* moments,

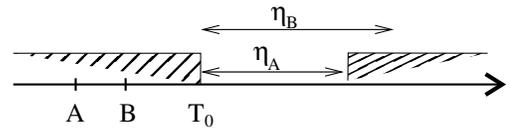

Fig. 2. Two stations, A and B start carrier sensing during some transmission. At time $T_0$ both wait, respectively, for $\eta_A, \eta_B$ and, detecting a free channel - commence their transmission. In this example, station A transmits.

rather than transmit with probability $p$ or back off for period $\lambda$.

*1) Two transmission points:* In this particular case, all waiting stations decide to do either of the following:

1) transmit immediately;
2) wait for period $\lambda$ and transmit;
3) attempt transmission at different time.

Let $p$ be the probability of immediate transmission and $q$ - the probability of transmission after period $\lambda$. Then, with probability $1 - (p + q)$ the station decides not to transmit at all. Note, that the probability that only one station transmits depends on $N$ (number of stations), $p$ and $q$ and is the following:

$$\Pr[\text{success}] = Np(1-p)^{N-1} + Nq(1-(p+q))^{N-1} \ . \quad (1)$$

We can determine what $p$ and $q$ maximize the success probability of the modified protocol. Let $f(p,q) = Np(1-p)^{N-1} + Nq(1-(p+q))^{N-1}$. One can show, that $f$ reaches maximum for $p = 1 - qN$ and

$$q = \frac{(N-1)^2}{N^2 \left(N - 1 - \left(\frac{N-1}{N}\right)^N\right)} \ .$$

Substituting to Eq. 1 we get for $N = 1, \ldots, 10$ success probabilities as follows: 0.666667, 0.612476, 0.589383, 0.576551, 0.568379, 0.562717, 0.558561, 0.555382, 0.55287.

*2) General case:* Assume there are $k \geq 1$ transmission moments to choose from (i.e. the transmission can start at either of the meoments $\lambda, 2\lambda, \ldots, k\lambda$). Then the probability of a successful transmission for a single station is

$$\Pr[\text{success}_{p_1,\ldots,p_k}] = \sum_{i=1}^{k} Np_i \left(1 - (p_1 + \ldots + p_i)\right)^{N-1}$$

and the overal time is $T_k = k\lambda + \delta$. We assume that $N \gg 1$ and try to find an optimal probability in the form $p_i = a_i/N$. Let

$$f_k(a_1, \ldots, a_k) = \sum_{i=1}^{k} a_i e^{-(a_1+\ldots+a_i)} \ .$$

Then

$$\Pr[\text{success}_{a_1/N,\ldots,a_k/N}] \sim f_k(a_1, \ldots, a_k) \ .$$

Let $(M_k)_{k \geq 1}$ be a series of real numbers defined by the following recurrence rule:

$$\begin{cases} M_1 = \frac{1}{e} \\ M_{k+1} = e^{-1+M_k} & \text{for } k \geq 1 \end{cases} \quad (2)$$

**Lemma 2** ([14]). *Function $f_k$ reaches maximum for $(b_k, \ldots, b_1)$, where $b_1 = 1$ and $b_a = 1 - M_{a-1}$ dla $a = 2, \ldots, k$ and this maximum is $M_k$.*

*C. Nonuniform continuous probability distributions*

We consider a variant of Cai, Lu and Wang algorithm where stations use a non-uniform distribution to choose backoff values within IP; throughout the rest of this section $f$ will denote density of this probability distribution.

A transmission in a slot succeeds, if $X_{(2)} - X_{(1)} > \delta$, where $X_{(1)}, \ldots, X_{(m)}$ denote the backoff values generated by stations after sorting them. Let $P_f(\delta, n)$ be the probability of this event. The following well-known theorem for the properties of order statistics differences was also formulated in [12]:

**Theorem 3.** *Let $n$ stations generate backoff values $(X_i)_{i=1,\ldots,n}$ using independently the density probability function $f(x)$ concentrated on the interval $[0,1]$. Let $F(t) = \int_0^t f(x)dx$ be the cumulative distribution function. Then*

$$P_f(\delta, n) = n \int_0^{1-\delta} f(x)(1 - F(x+\delta))^{n-1} dx \ , \quad (3)$$

*Proof:* First recall some simple facts. Suppose that $X_1, X_2, \ldots, X_n$ are independent random variables with pdf $f(x)$ and corresponding cdf $F(x)$. Let $X_{(1)}, X_{(2)}, \ldots, X_{(n)}$ be ordered statistic. Then, the joint pdf of $X_{(1)}$ and $X_{(2)}$ is given by

$$f_{12}(x,y) = n(n-1)f(x)f(y)[1-F(y)]^{n-2} \ .$$

Therefore

$$P(\delta, n) = \int_0^{1-\delta} \int_{x+\delta}^1 f_{12}(x,y) dy dx = $$
$$n \int_0^{1-\delta} f(x)(1 - F(x+\delta))^{n-1} dx \ .$$

□  ∎

After some calculations (see [12] once again for details) we obtain that with a probability at least $1 - \frac{1}{n}$ successful leader election is at most

$$\frac{1 + \frac{\lambda}{\delta}}{P_f(\delta, U)} \cdot n + O(\sqrt{n \log n}) \quad (4)$$

where $U$ is a fixed upper bound on the number of stations. Cai, Lu and Wang considered the uniform probability on the interval $[0,1]$ (in this case we have $P_{unif}(\delta, n) = (1-\delta)^n$) and probabilities with the cumulative distribution function of the form $F(t) = t^{r+1}$, where $r \geq 0$.

We propose to use distributions. Namely, let $f_{\alpha,\beta}$ be the probability density function concentrated on the interval $[0,1]$ with the cumulative distribution function

$$F_{\alpha,\beta}(x) = \frac{e^{\alpha x^\beta} - 1}{e^\alpha - 1},$$

where $x \in [0,1]$ and $\alpha, \beta > 0$. One can check that $f_{\alpha,\beta}(x) = \frac{\alpha \beta x^{\beta-1} e^{\alpha x^\beta}}{e^\alpha - 1}$. We show the following theorem:

**Theorem 4.** *Let $\alpha, \delta > 0$ and $n$ be a natural number. Then*

$$P_{f_{\alpha,1}}(\delta, n) = e^{-\alpha \delta} \left(\frac{e^\alpha - e^{\alpha \delta}}{e^\alpha - 1}\right)^n \ . \quad (5)$$

In the case when $\beta = 1$ we used the last equality to determine for given $\lambda$ and $U$ the optimal parameters $\alpha$, $\delta$. For $\beta \geq 2$ we solved the problem

$$(\alpha^*, \delta^*) = \arg\min_{\alpha, \delta} \frac{1 + \frac{\lambda}{\delta}}{P_{f_{\alpha,\beta}}(\delta, U)}$$

## V. SIMULATIONS

We performed a range of simulations of the proposed changes, testing the throughput under different network ($N$ - number of nodes, $\lambda$ - duration of a single packet transmission)) and algorithm parameters ($k$ - number of possible transmission points). We modelled the number of packets generated in the network using two approaches:

FB  (full buffer) – at each moment, each station has a packet ready to be sent; after successfull transmission a new packet is available immediately;

POS (Poisson) – new packets are generated according to the Poisson process.

The network is single-hop and no delays due to distance between nodes or fading phenomena were modelled. The packet duration is $\delta = n\lambda$ for $n \in 1, \ldots$, and the base time-unit is $1\lambda$. Figure 3 presents an example run of the algorithm.

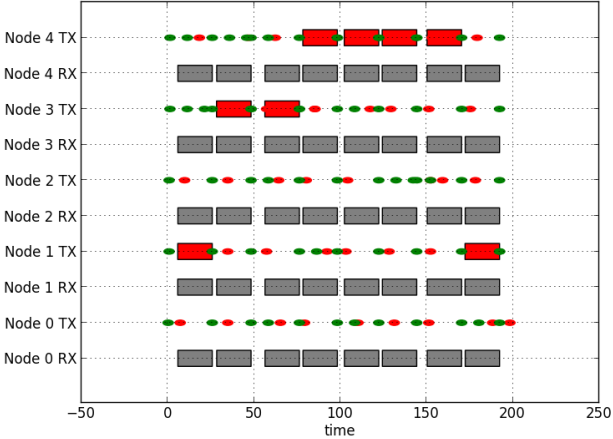

Fig. 3. Timings for a single simulation run for $N = 5, k = 1, \delta = 20\lambda$, FB data generation. ■ - transmission periods, ■ - carrier can be sensed, ● - choosing transmission point, ● - other transmission detected

Simulation data was obtained for the following parameters:
- network size $N = 5$;
- packet length $\delta = 100\lambda$;
- variable number of transmission points $k = 1, \ldots, 15$;
- packet generation POS with average interlude between packets $= 100\lambda$ and FB.

In order to asess the behaviour of the algorithm, the following metrics were measured:
- delay of a packet, $T_{delay}$, which is a time interval between the moment the packet is generated by a station and the moment of its successful transmission;
- channel occupancy, $T_{busy}$, measured as the total simulation time less the sum of all times when no station transmits (to transmission because all stations wait for their periods $\lambda$);
- system throughput, measured as the number of packets transmitted $N_{TX}$ and received without collision $N_{RX}$.

In Fig. 4 the throughput of the network for both packet generation scenarios is presented. As $k$ increases, which yields increasing the period during which the stations compete, there are fewer transmission attempts in the same simulation period. Simultanously, the number of correctly received packets increases, because the transmissions are less likely to collide. On balance, the proposed modification improve overal throughput of the network as well as roubustness of communication.

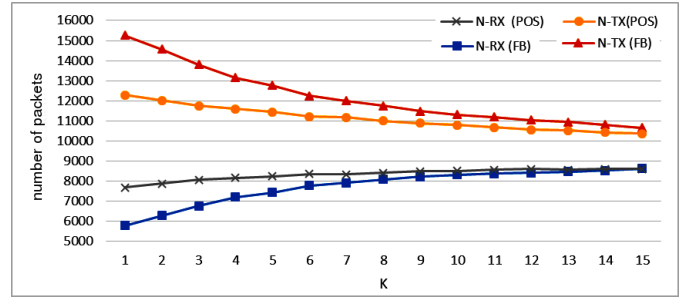

Fig. 4. Number of packets sent ($N_{TX}$) and correctly received ($N_{RX}$) as function of $k$.

Channel occupancy for both packet generation scenarios is presented in Fig. 5. The time during which the channel is free is the time spent on delays incurred by leader election process. We determined this time to be between $\approx 1\%$ and $\approx 6\%$ of the overall time. The fact that channel occupancy is smaller in POS scenario is explained by the delays in packet generation – that means that the channel can accomodate more packets than the underlying simulation process could provide.

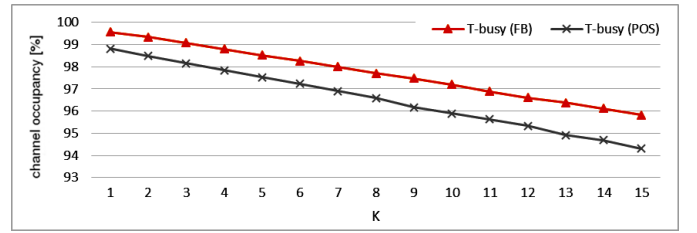

Fig. 5. Channel occupancy during simulation period.

An important efficiency factor for medium access protocols is the delay incurred by the protocol to the packet transmission ($T_{delay}$ defined above). In Fig. 6 we show average packet delay for given scenario. Clearly, smaller delays are observed for POS generation of packets, which corresponds to the smaller channel occupancy as depicted above.

Lastly, based on experimental data for attempted and successful transmissions, it was possible to determine the probability for successful transmisssion as a function of $k$. This is plotted in Fig. 7 along with theoretical probability as given in

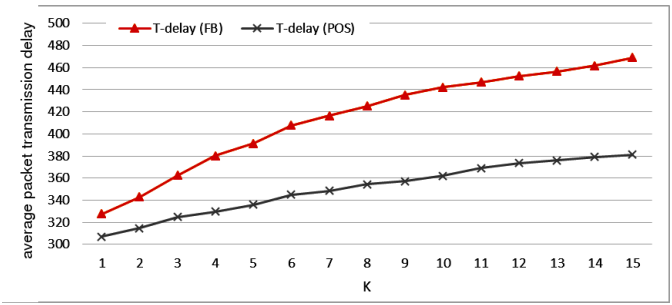

Fig. 6. rednie opnienie transmisji pakietu.

Eq. 2. Clearly, the real (from simulation) probability is better than calculated, but both values seem to converge as $k$ grows bigger.

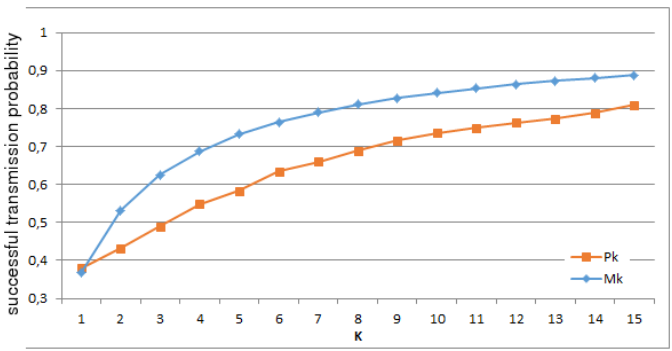

Fig. 7. Comparison of probability of a successful transmision determined theoretically ($M_k$) and by experiment ($P_k$) as funciton of $k$.

Considering these results, one can conclude that while $k$ gets bigger, the reliability of transmission increases, at the cost of packet delay. Notably, between $k = 1$ and $k = 15$, the delay grows by one length of the packet (cf. Fig. 6). At the same time, the channel utilisation drops, as more time is spent on avoiding collision.

## VI. CONCLUSIONS

Presented modifications for solving medium access problem in ad-hoc wireless networks have better properties than simple approaches considered before (for example, case from paper [13] corresponds to our solution for $k = 1$).

Asymptotic properties of the resulting protocol are confirmed by simulations. The probabilities of successful (non-coliding) transmission ($M_k$) correspond to analogous obtained in simulator ($P_k$). We have shown that the idea of introducing more than one transmission moment ($k$) brings considerable profit in terms of success rate – for the same number of sent messages, there is 50% more successfuly received messages when $k = 15$ than when $k = 1$. However, this comes at the cost of longer delays of messages (times between message creation and its successfil receive), and lowered channel utilisation.